\documentclass[a4paper]{article}
\usepackage{palatino}
\usepackage{helvet}
\usepackage{courier}
\usepackage{url}
\makeatletter
\usepackage{graphicx}
\usepackage{color}
\usepackage{soul}
\usepackage{booktabs}
\usepackage{longtable}
\usepackage{hyperref}

\usepackage{tabularx}

\usepackage[dvipsnames]{xcolor}
\usepackage[most]{tcolorbox}
\definecolor{block-gray}{gray}{0.87}
\newtcolorbox{citationblock}{colback=block-gray,
boxrule=0pt,boxsep=0pt,breakable}

\usepackage{subcaption}
\usepackage{amsmath}
\usepackage{xcolor}
\usepackage{graphicx}
\usepackage{verbatim}
\usepackage[inline]{enumitem}
\usepackage{booktabs}
\usepackage{todonotes}
\usepackage[normalem]{ulem}

\usepackage{caption}
\usepackage{tabularx}
\usepackage{fancyhdr}
\usepackage{hyperref}
\usepackage{authblk}
\usepackage{hyphenat}

\newcommand{\aquote}[2]{\textit{\normalsize{#2: "#1"}}} 
\newcommand{\aquotetab}[2]{\textit{\normalsize{#2: "#1"}}}

\date{}

\renewenvironment{abstract}
 {\small
  \begin{center}
  \bfseries \abstractname\vspace{-.em}\vspace{0pt}
  \end{center}
  \list{}{
    \setlength{\leftmargin}{30pt}%
    \setlength{\rightmargin}{\leftmargin}%
  }%
  \item\relax}
 {\endlist}

\def\platforms{music platforms}
\def\platform{music platform}

\begin{document}

\title{What is fair? Exploring the artists' perspective on the fairness of music streaming platforms}
%
%
%

\author[1]{Andrés Ferraro}
\author[1]{Xavier Serra}
\author[2]{Christine Bauer}

\affil[1]{Universitat Pompeu Fabra, andres.ferraro@upf.edu}
\affil[2]{Utrecht University, c.bauer@uu.nl}
%
\maketitle     
\thispagestyle{empty}

\begin{citationblock}
\flushleft
This is the accepted version of the paper appearing in the Proceedings of the 18th IFIP International Conference on Human-Computer Interaction (INTERACT 2021). 
%

\vspace{10pt}

\textit{Andrés Ferraro, Xavier Serra and Christine Bauer. ``What is fair? Exploring the artists' perspective on the fairness of music streaming platform''. 18th IFIP International Conference on Human-Computer Interaction (INTERACT 2021). Bari, Italy.
}

\end{citationblock}
\vspace{20pt}

\begin{abstract}
Music streaming platforms are currently among the main sources of music consumption, and the embedded recommender systems significantly influence what the users consume. 
There is an increasing interest 
to ensure that those platforms and systems 
are fair. Yet, 
we first need to 
understand what fairness means in such a context. 
Although artists 
are the main content providers 
for music 
platforms,
there is a research gap concerning the artists' perspective. 
To fill this 
gap, we conducted 
interviews with music artists to understand how they are affected by current platforms and what improvements they deem necessary. 
Using a Qualitative Content Analysis, we identify the 
aspects that the artists consider relevant for fair platforms. 
In this paper, we discuss the following aspects derived from the interviews: 
fragmented presentation, reaching an audience, transparency, influencing users' listening behavior, popularity bias, artists' repertoire size, quotas for local music, gender balance, 
and new music. 
For some topics, our findings do not indicate 
a clear direction 
about the best way how 
music platforms should act and function; 
for other topics, though, there is a clear consensus among our interviewees: for these, 
the artists have a clear idea of the 
actions that should 
be taken so that music platforms will be
fair 
also 
for the artists.

\end{abstract}
\section{Introduction}\label{sec:introduction}
Music streaming platforms are currently among the main sources of music consumption. What the users consume is strongly influenced by what is offered on the \platforms{}, and what is promoted through the algorithmic recommendations in particular. What the users consume, in turn, shapes the music streaming ecosystem at large. 
There is an increasing research interest in making the platforms and their recommendations fairer. 

Generally, the 
topic of fairness received 
particular attention in machine learning and artificial intelligence~\cite{Hutchinson2019_unfairness}, focusing on the quantification of fairness and the algorithmic perspective.
Research in the field of human-computer interaction (HCI), in contrast, studied the perception of fairness by taking the end consumers' perspective (e.g., \cite{HELBERGER2020105456,harrison2020,wang2020}). 
In general, a large part of HCI research addresses the user who directly interacts with a (computational) system. Yet, also other humans are impacted by system design and HCI practices. Hence, it is crucial to consider also these groups' demands in design considerations.

Recommender systems connect and affect several stakeholders including consumers, the platform provider, item providers, and society~\cite{jannach2020_mcnamara}.
While several potential sources of bias have been identified 
that may lead to unfairness (e.g., \cite{yao2017beyond,marlin2012collaborative,farnadi2018fairness}
), the focus of attention lies on algorithmic and data bias~\cite{baeza2016data}. 
A recent review~\cite{Milano2020} highlights that existing literature tends to consider the implications for the receivers of the recommendations (i.e., end consumers) only. Thus, the research gap lies in considering the provider's interests (and the interests of the society at large) when assessing the ethical impact of recommender systems. 

Addressing this research gap, this work focuses on the item provider's perspective on online platforms, which are an underrepresented group of humans in HCI research. 
%
This also holds for the music domain. 
Recommender systems play a prominent 
role in connecting users with content (and the artists behind this content) on such \platforms{}.
While various studies (e.g., \cite{bauer2019_PLOSONE,kowald2021support}) 
have investigated and shown that current music recommender systems (MRS) do not serve different user segments equally well, there is a research gap concerning 
the artists' perspective~\cite{bauer2019,sigirforum2020_helpartists}. 
Besides being a so-far underrepresented group of humans in HCI research, music artists have limited opportunities for direct interaction with a \platform{} (in their role as an item provider)\footnote{
Interaction is limited to providing 
recordings and some meta-data (e.g., title, tags).}; and yet, they are strongly affected by the design of such systems. 
%
%
While our long-term goal is to have \platforms{} that are perceived `fair' from the various stakeholders' perspectives, this work 
zooms in music artists as the main item providers for \platforms{}. 
The concrete goal of this work at hand is to \emph{understand the different dimensions of \platforms{} (and their recommender systems) that define fairness from the artists' perspective}. 


To this end, we conducted semi-structured interviews with 
music artists from different countries and with varying popularity. 
Using a Qualitative Content Analysis as proposed by 
\cite{mayring2004qualitative}, we have (i)~identified various 
aspects that show how the artists feel affected by the current \platforms{} and their integrated information retrieval and recommender systems components, and (ii)~how such \platforms{} and systems could be improved concerning 
fairness. Only such understanding can ultimately lead to fairer \platforms{} that are not only optimized for the interests of the platform providers or the consumers. 
%
%
We deem our work a fundamental contribution to (i)~understanding the impact that such \platforms{} have on artists from their perspective and (ii)~understanding what the artists consider fair for them. 


%

The remainder of this work is structured as follows: First, we present the conceptual basis and discuss related work (Section~\ref{sec:related}). In Section~\ref{sec:methods}, we detail the methods. Section~\ref{sec:results} represents the major part of this paper, where we outline and discuss the issues and aspects 
that we derived from the interviews. 
As we illustrate the artists' concerns and ideas, we 
embody a completely novel contribution to the field.
In Section~\ref{sec:discussion}, we discuss how the findings reflect 
the existing understanding of fairness and derive implications for the operationalization of recommender systems and interaction functionalities on \platforms{}. 
In Section~\ref{sec:conclusion}, we summarize our work and point to future research directions.

\section{Conceptual Basis and Related Work}\label{sec:related}



A purely technical approach is not sufficient for defining and operationalizing fairness in practice, as seen in previous attempts in several fields~\cite{cramer2018assessing,cramer2019translation,Madaio2020,Holstein2019}. 
Taking an information systems perspective, \cite{Feuerriegel2020} emphasize that fair systems need to embrace people, technology, and organizations, and that unsolved challenges exist on all three dimensions.
In other words, fairness cannot be operationalized with a lack of a definition or understanding of what fairness connotes in a certain context.
The notion of fairness has evolved over time~\cite{Hutchinson2019_unfairness}; and particularly recent literature (e.g.,~\cite{Holstein2019,selbst2019}) emphasizes that for developing a good understanding of fairness in a given context, it is crucial to 
take a multidisciplinary point of view and listen to the opinions of the 
various stakeholders involved and
affected. 

\subsection{Defining Algorithmic Fairness
}\label{sub:alg}


There 
is a multitude of 
definitions of (algorithmic) fairness 
\cite{Hutchinson2019_unfairness}. 
Two common definitions are
`group' and `individual' fairness. Individual fairness reflects 
that similar individuals 
should be treated similarly. Group fairness ensures that people of a protected 
group should be treated in the same way as the rest of the population. 
While 
\cite{dwork2012fairness} clearly distinguishes those two concepts, 
other works (e.g., \cite{binns2020apparent}) suggest that individual and group fairness 
are not contradictory and may be achieved simultaneously. 
In 
information retrieval, 
fairness is frequently defined in terms of \emph{exposure}~\cite{biega2018equity,sapiezynski2019quantifying} or \emph{attention}~\cite{singh2018fairness}.
Thereby, 
\cite{biega2018equity} focus on individual fairness, whereas other works typically consider group fairness~\cite{singh2018fairness,sapiezynski2019quantifying}. 
The idea of exposure of the artists could also be adopted for 
the music domain. Yet, it is not answered how a \emph{fair exposure} should be operationalized or---on an even deeper level---what fairness actually means in the context of \platforms{}. To understand what fairness is from the artists' perspective, we need to 
involve artists. 
To the best of our knowledge, there is no public work that reaches out to artists to identify how they feel affected by current \platforms{} and how they believe the embedded recommenders systems could be improved to be fair for them.

\subsection{Perception of Fairness in Algorithms}\label{sec:perception}
Besides 
defining algorithmic fairness, there is a wealth of studies investigating how people perceive and reason about the fairness of algorithms. 
When it comes to definition, people 
seem to prefer 
simple 
compared to complex ones~\cite{srivastava2019}.

In a survey investigating the general perception of fairness of automated decision-making systems, \cite{HELBERGER2020105456} 
found that people 
consider that systems 
make more objective decisions than humans and may also process larger amounts of information 
which would allow them to make fairer decisions. Yet, the 
respondents also noted that 
systems are limited concerning 
generalizability and modeling 
reality. Another survey~\cite{harrison2020} studied how people perceive the fairness of 
realistically-imperfect systems. 
As the respondents had contradictory opinions on 
systems, the authors conclude 
that is impossible to achieve a broad acceptance in society regarding the ``right'' fairness definition. The study also highlights that there is a general preference towards human judges compared to a system, even if the participants consider the systems fair or unbiased.

In an online experiment \cite{wang2020}, 
participants rated algorithms according to their perception of fairness. The study found that participants rate systems as more fair if they are favored by it, 
even in cases where the algorithms were explicitly described to the participants as being very biased to a particular demographic group. This effect was present across various participant groups, in different levels 
depending on education level, gender, and other aspects of the participants. 

In group discussions and interviews with 
traditionally marginalized groups in the US, \cite{woodruff2018} sought to understand these groups' 
perceptions of fairness in algorithms. This work highlights that the opinions 
regarding the fairness of algorithms vary depending on individual factors, context, different stakeholders' perspectives, and different framing of fairness. This can help to explain inconsistencies in some findings across studies on fairness perception. 
Further, this shows that the contextual factors should be taken into account when studying algorithmic fairness, supporting the idea of considering the interests of the different groups of stakeholders of the systems.



\subsection{Bias and Fairness in Music Recommendation}\label{sec:fair_mrs}

Besides the general issues and considerations concerning fairness, the \platforms{} and their recommenders face specific challenges. Among others, the inherent popularity bias is widely discussed. In this section, we discuss related works that study how music recommender can affect artists. 

The music market is a typical example of the long-tail economy~\cite{RN26,RN27} with a highly uneven distribution of demand for the most popular and the least popular items. Since Anderson~\cite{RN26,RN27} popularized the concept, 
there is an ongoing debate about whether 
online 
platforms 
indeed facilitate users to 
consume items 
in the long tail 
or, on the contrary, 
accentuate the `superstar phenomenon'~\cite{rosen1981} where the popularity curve gets more skewed in the long-term~\cite{coelho2019digital,bauer2017music,ferraro2020exploring}. 
%
%
Celma~\cite{celma2010music} was among 
the first to study 
how 
different recommenders 
may or may not promote 
the less popular items in the music domain. 
Also for other 
domains, \cite{fleder2009blockbuster} 
found that 
users tend to consume a reduced number of items when recommender systems are at play. 
While these works focused on the consumers' behavior, the findings clearly indicate that recommenders have an impact on item providers.

A study in the Spotify context~\cite{anderson2020algorithmic} found that users who follow 
algorithmically-generated recommendations have 
reduced diversity in the content they consume. 
In contrast, users who consume increasingly more diverse content, are those who reduce the algorithmic consumption and increase their organic consumption. 
As reducing the diversity in recommendations means that fewer artists get exposure 
and that it is more difficult for them to reach an audience, we believe that is important to also study 
the effect that algorithmically-generated recommendations can have on artists (e.g.,~\cite{ferraro2019,ferraro2020artist}.) 

Based on the idea that many local artists tend to be obscure long-tail artists, there are research endeavors (e.g., \cite{turnbull2018local,akimchuk2019evaluating}) to ``localify'' recommendations 
to promote local artists. Although 
technical approaches to localize music recommendations are proposed, 
the artists' perspective on 
such a goal or approach is not discussed. 
Another work by Spotify~\cite{way2020local} studies music consumption on a country level. 
The latter study investigates how users from a country consume content from another country, and how this consumption pattern evolves and changes over time. The authors identify that language and geographical proximity potentially 
impact 
the consumption between countries. 
Yet, a more fine-grained investigation is needed where 
also the popularity of the artist is taken into account. Such an endeavor may allow drawing conclusions about the chances given to local artists for gaining a larger audience using these platforms. 

Mehrotra et al.~\cite{mehrotra2018towards} acknowledge that fair music recommenders need to consider multiple stakeholders. They define the recommendation problem 
as a marketplace and 
consider the perspectives 
of the music consumers and the artists. Intending 
to achieve algorithmic fairness, they define a fairness metric for 
artists that they base 
on the popularity distribution in the recommendations. 
While that work points out 
that there are different ways of defining fairness, 
a rationale for their choice and definition are missing.


However, previous work rarely reaches out to artists and, to date, no work involves music artists for understanding how \platforms{} can be 
fair for them. Baym~\cite{baym2018playing} builds on interviews with artists to investigate how digital technologies changed their interaction with the audience. 
Andersen and Knees~\cite{andersen2016conversations} interview artists for investigating their assessment of algorithms and artificial intelligence in the music creation process; our work, in contrast, refers to recommenders for music consumers. 
Aguiar and colleagues~\cite{aguiar2018platforms,aguiar2017let} analyze how strategic decisions and recommenders embedded in \platforms{} 
affect music consumption behavior and 
the success of songs. 
Ferraro and colleagues~\cite{ferraro2019,ferraro2020artist} 
take an algorithmic approach to analyze the impact of MRS on users' behavior 
and how this 
affects 
different groups of artists. 
Holzapfel et al.~\cite{holzapfel2018ethical} raise 
ethical implications in music information retrieval and conceptually illustrate how \platforms{} may negatively impact certain groups of artists. 

\section{Methods}
\label{sec:methods}
To understand the artists' perspective on the current \platforms{} and their embedded MRS, we conducted 
semi-structured interviews with music artists. The interviews took place from December~2019 through March~2020. 
The interviews are part of our ongoing research 
to explore viable solutions for fairer \platforms{} and MRS, 
this work at hand focuses on understanding the artists' perspective on what represents a fair \platform{}. 

We carried out $9$ semi-structured interviews with music artists. 
According to research practice~\cite{creswell2016qualitative,morse1994}, the sample size is adequate and, more importantly, we reached a high level of thematic saturation~\cite{Guest2006_saturation} with the same topics being repeatedly mentioned across the interviews.
The interviews were designed to last one hour in total for each interview including a 
brief and general introduction to MRS and \platforms{}. We used open questions with the aspiration 
that the artists would bring up their own 
ideas that might not be considered in the field before. 
In addition, we proposed 
specific alternatives to gather the artists' opinions on those alternatives.

In the following, we detail the interview process. The materials used for the interviews (i.e., invitation letter, consent form, first version of questions, final version of questions) can be found on Zenodo\footnote{\url{https://doi.org/10.5281/zenodo.4793395}}.

\subsection{Interviews}
Before the interviews, we informed that the data and results of the interviews will be kept anonymous at all times, which was important so the artists could feel free to share concerns regarding the \platforms{} or issues related to the music industry. 
In addition to the consent form, we asked the participants to fill out a short form with optional information about themselves that would be used to refer to their answers (i.e., age, gender, country, genre, popularity level, number of records/singles, 
years active in the music industry, contracts with labels).
The first 10~minutes of the interview were used to explain the project's purpose and to give a general introduction to MRS and how these are integrated into current \platforms{}. 

For the interview part, we defined a 
protocol to be used during the interviews which included a set of guiding 
questions and tentative question to encourage the interviewees to elaborate further. The protocol was developed so that each guiding question addressed and explored various topics and issues of how the \platforms{} might affect artists. As the first step in elaborating this guide, we started with collecting information and ideas about the general aspects of current \platforms{} that could affect artists. In a second step, we formulated questions to address the identified issues. 
Since the rich collection of potentially interesting issues had to be reduced to fit the time scope of an interview and to have (a narrower) focus, in the third step, each team member gave a score between 1 and 3 to each of the original 36~questions according to priority, with 1 being the highest priority. Afterward, the team discussed the 14~questions with differing scores until consensus was reached.
Based on the agreement between the scores we defined a definitive list of guiding 21~questions (whereof we use 11~as main questions and 10~as sub-questions) to be used in the interview protocol (Table~\ref{tab:questions}). 
Note, all interviews were held in Spanish and the materials provided to participants were all in Spanish.

\begin{table*}[!h]
  \centering
  \caption{Guiding questions in the interview protocol.} 
  \label{tab:questions}
  \begin{tabularx}{\textwidth}%
    {l>{\setlength\hsize{0.19\textwidth}\raggedright}X X}
  \toprule
  No. 
  &
  Topic&Guiding question \\
  \midrule
  1&Convey interest, gain trust&Do you use any platform to listen to music? What's your experience with it?\\

  2&Reflecting&Do you think your career would be much different without these systems? \\

  3&Lack of control&Which of your music tracks should be recommended more and which less?\\

  4&Bias to more popular&There are groups of artists that are not recommended by the system because of different reasons. Do you see any alternatives for this?\\

  5&Diversity&Music considered ``niche music'' is not recommended to many users, should the system nurture diversity (e.g., in terms of genres, styles, artists from all over the world, popular and not-yet-popular) or focus more on recommending what the user is familiar with?\\

 6&Size of repertoire&If artist X has more music than the artists Y, do you think the system should recommend more music by artist X---or should the recommendation be independent of an artist's repertoire?\\

 7&New artists&For a given user out of 100 recommendations, how many do you think should be new artists?\\

8&New music&Should your older songs be promoted more than your newer songs?\\

9&Country quotas&In a \platform{} that has more users from country X but more artists from country Y, the artists from X could be recommended more than artists from Y. What is the behavior that you expect from the system in that case?\\

10&Influencing the users&Currently, K-pop is the 7th most listened genre, over R\&B and classical music. Such a popularity distribution could also refer to gender, country, or other aspects of the music. Do you think the systems should try to reproduce this behavior, or should try to provoke a change on it? \\

11&Income distribution&Do you think the current model based on number of streams is good or there could be a better model?\\

  \bottomrule
  \end{tabularx}
\end{table*}

We note 
limitations of our interview design. 
First, while we cater for diversity, our sample is not 
representative of the population of music artists. The findings we report should, thus, be viewed as a deep exploration of our sample's beliefs and attitudes, but not as generalizing to the 
artist population as a whole. 
Yet, the findings indicate that we reached a high level of saturation with the same topics being repeatedly mentioned across the interviews. 
Second, while we assure anonymity in the presentation of the results, 
the interviewer naturally knows the participants' identity, which may have influenced participants such that some issues may not have been voiced. As some of the participants sometimes used strong language and addressed delicate issues, we believe that we were able to maintain a comfortable atmosphere with a high level of trust. 

\subsection{Participants}

We recruited $9$ music artists that we consider diverse---in the kind of music they perform, their popularity, location, age, and gender. Four of the participants are between 26--35 years old, four are between 36--45, and one is within the range 46--55. Seven are male and two are female. Most of these artists have many projects in parallel (one is a solo artist, the others work with many bands). 
The artists were born in four different countries (i.e., Australia, Spain, Russia, Uruguay) and started their careers in three different countries (i.e., Spain, Uruguay, Cuba).

The genres that the artists consider their music sum up to a total of 24. 
Examples are folk, pop, ska, punk-rock, dubstep, D\&B, and world music. The number of years in the industry is between 4 and 25. The number of albums released ranges between one and ten. Five participants consider themselves `independent artist', three have a contract with one of the three major labels (i.e., Sony, Universal, Warner), 
two have a contract with an independent label. 
Five artists consider themselves internationally known, 
four are known within their country.
The specific information of each participant is given in Table~\ref{tab:part}, including the identifier that we use to refer to participants for quotes.

\begin{table*}[ht]
  \centering
  \caption{Information about the participants.}
  \label{tab:part}
  \footnotesize
  \begin{tabularx}{\textwidth}{l l c >{\raggedright}X l l l}
  \toprule
  ID & Country&  Age 
  & Music Styles & Audience & Gender 
  & Contract \\
  \midrule
  P1 & Uruguay& 46-55& Rock, Folk, Hip-Hop, Electronic&International& Male&Major Label\\
  P2 & Uruguay& 26-35& Rock, Hip-Hop, Reggae, Dub& Local&Male&Major Label\\
  P3 & Uruguay& 36-45& Ska, Punk/Rock, Dub, Dubstep, D\&B&International&Male &Major Label\\
  P4 & Spain& 26-35& Indie, Rock& Local &Male&Independent\\
  PF1 & Cuba& 36-45& World, Jazz, Cuban Music, Electronic&International&Female&Independent\\
  PF2& Spain& 26-35& Indie Pop, Singer-songwriter& Local &Female&Indie Label\\
  PN1 & Uruguay& 26-35& Alternative Rock/Indie, Progressive Rock& Local&Male& Independent\\
  PJ1 & Spain &36-45& Jazz, Free Improvisation&International & Male& Independent\\
  PR1& Spain & 36-45 & Hip-Hop, Reggae, Blues, Salsa, Flamenco & International &Male&Indie Label\\
  \bottomrule
  \end{tabularx}
\end{table*}

\begin{table*}[!h]
  \centering
  \caption{Details on the annotation scheme.} 
  \label{tab:annotationscheme}
\begin{tabularx}{\textwidth}{>{\setlength\hsize{0.133\textwidth}\raggedright}X >{\setlength\hsize{0.419\textwidth}\raggedright}X X}
  \toprule
  Topic&Description&Example of annotation \\
  \midrule
  User view & Participant comments on using a \platform{} in the role of a music consumer.
  &\aquotetab{I usually read the artist's biography and the influences of an artist.}{PN1}\\

  Artist view&Participant expresses opinion from the point of view of their artist role.&\aquotetab{[...] it wouldn't hurt if the systems were 
  more random, not that obvious---if you like `Beatles,' I recommend 'Rolling Stones.'}{P2}\\
  
  Lack of control&Reference to giving more control either to the artists over the music presented, or to users over what they get recommended.& \aquotetab{As an artist, I would love to have more freedom of action over my music on the platform.}{P2} 
  \\
  
  Diversity&Related to any aspect of diversity in the recommendation.& \aquotetab{
  I would expect that the \platforms{} promote more diverse content.}{P3} 
  \\

  Context of music&Aspects related to information and presentation apart from the music; 
  the context that it is embedded in. 
  &\aquotetab{There are songs that have history, [you cannot ignore that].}{P1}\\

  New music&Participant refers to artists new on a \platform{} or new music of existing artists.&\aquotetab{
  [...] it makes sense if half of the recommendations [made by the \platform{} are songs of] new artists.}{P2} 
  \\

  Popularity bias&Participant refers to aspects related to popularity bias of recommender systems or the music business.&\aquotetab{The problem is [that] the recommender system systematically ignores all those potential artists because it is easier to recommend [what is more popular].}{P4}\\

  Influencing 
  behavior/ taste&Participant expresses opinion concerning \platforms{}' opportunity to influence users'
  listening behavior or music taste.&\aquotetab{In my opinion, you can't impose some [specific] 
  music to the users.}{PF1}\\
 
Transparency&Refers to the need for information about how a \platform{} works and how its recommender system makes decisions.& \aquotetab{If a human makes that decision if he says `I have a small store, I am going to put it this way,' I will understand it better than if an algorithm does it.}{PN1}\\

Labels'/ platforms' interests&Participant refers to the interests of stakeholders such as the \platform{} providers or record companies.&\aquotetab{[A platform] needs to take responsibility for its recommender system 
[...] Obviously, they are commercially not capable or not interested.}{P2}\\
Size of artists' repertoire&Participant distinguishes between artists with more or fewer songs or albums.&\aquotetab{If you have more songs, you have more chances to satisfy different audiences.}{PR1}\\
Quotas for local music&Participant mentions regulations for local music such as 
quotas for local artists on \platforms{}.&\aquotetab{Otherwise, you won't know what there is in your country.}{PR1}\\
Gender balance&Participant talks about gender bias in the music industry or on \platforms{}, or how recommendations might be fair(er) from a gender perspective.
&\aquotetab{[...] the population of the world is 50\% women. So it would be ridiculous if the system wouldn't recommend it.}{PN1}\\
Regulation of recommendations&Participant refers to regulations or policies for \platforms{} or their recommender systems.&\aquotetab{[...] the question is if it should be imposed by the state [to promote local music].}{P3}\\
Royalties 
distribution&Participant refers to 
royalties generated on \platforms{} or their distribution 
among artists.&\aquotetab{It is absurd what [the platform] pays to the artists.}{PF1}\\
\bottomrule
\end{tabularx}
\end{table*}

\subsection{Processing and Analysis} 

Following the methodology of Qualitative Content Analysis~\cite{mayring2004qualitative}, the interviews were recorded and transcribed, followed by developing an annotation scheme (i.e., coding) 
inductively from the transcriptions of the interviews, which we used to annotate the transcriptions accordingly.


The total duration of the recordings is 420~minutes, which transcribed 
correspond to 33,669 words. 
The annotation scheme was developed inductively from the transcriptions, where statements were used as the level annotations. Often, statements were on sentence level; yet, many sentences include two or more statements. Note that the annotation scheme was developed in English while the transcriptions were kept in their original language (i.e., Spanish).
The development of the annotation scheme and the annotation of statements itself was an iterative process where we assigned a topic to a specific sentence and if it did not fit to any of the previous topics then a new one was defined. We iterate a total of 4~times and the final annotations were reviewed by a different person than the annotator 
to increase intercoder reliability. 

The final annotation scheme 
includes a total of 
15~overall topics that were 
obtained from 752~annotated text sections. 
In Table~\ref{tab:annotationscheme}, we describe 
the topics with an example 
annotation for each topic. Note, quotes are given as translations to English, whereas the interviews were held in Spanish.
Table~\ref{tab:stats} presents the number of annotations per participant per topic. 



\begin{table*}[!h]
  \centering
  \caption{Statistics about annotations.}
  \label{tab:stats}
  \begin{tabular}{lrrrrrrrrrrrrrrr|r}
  \toprule
  Topic & P4&PF1&P3&P2&P1&PN1&PJ1&PR1&PF2& Total\\
  \midrule
  User view &8& 1& 5& 8& 6& 5& 4& 2& 7&46\\
  Artist view&15& 12& 25& 22& 14& 17& 4& 9& 13&131\\
  Lack of control&15& 4& 8& 14& 6& 5& 0& 2& 8&62\\
  Diversity&5& 2& 8& 6& 4& 2& 1& 3& 8&39\\
  Context of music&7& 6& 8& 2& 17& 1& 0& 5& 0&46\\
  New music&12& 6& 7& 8& 6& 14& 6& 10& 13&82\\
  Popularity bias&6& 0& 8& 5& 1& 4& 2& 5& 1&32\\
  Influencing user behavior/taste&7& 2& 17& 7& 16& 3& 1& 7& 7&67\\
  Transparency&7& 0& 11& 13& 2& 6& 0& 2& 2&43\\
  Labels'/platforms' interests&8& 4& 14& 17& 13& 5& 5& 11& 5&82\\
  Size of artists' repertoire&2& 1& 2& 2& 2& 4& 1& 2& 1&17\\
  Quotas for local music&3& 3& 8& 3& 5& 5& 2& 2& 3&34\\
  Gender balance&5& 3& 5& 4& 0& 2& 1& 1& 2&23\\
  Regulations of recommendations&3& 0& 3& 0& 4& 0& 0& 0& 0&10\\
  Artists' income distribution&3& 1& 4& 9& 4& 4& 3& 7& 3&38\\
  \hline
  Total& 106&45&133&120&100&77&30&68&73&752\\
  \bottomrule
  \end{tabular}
\end{table*}

\section{
Results}\label{sec:results} 
In the first part of this section (Sections~\ref{subsec:fragmentary} to \ref{subsec:transparency}), we present how the artists feel affected by current \platforms{}. Subsequently (Sections~\ref{subsec:influencing} to \ref{subsec:newmusic}), we discuss those topics that give direction on what the artists consider fair \platforms{}; in the paper at hand, we report those topics that were addressed by \emph{all} participants. Table~\ref{tab:concl} provides an overview of these topics and summarizes what the participating artists deem necessary for future fair MRS. 

\subsection{Fragmented Presentation}\label{subsec:fragmentary}

The participants report that they do not find adequate the way they are presented on the \platforms{}. 

Two artists (P2 and P3) mention that their artist profiles on the \platforms{} show old tracks at the top 
because those are the most-listened items over the years. \aquote{But it is something that I have done 10~years ago. [The platform] puts the most listened tracks. [...] 
and you have to scroll down to reach the latest album.}{P3} 

Also, the context in which artists and their music are presented may affect their public image; e.g., the artist P3 refers to a feature that 
serves users with an automatically generated playlist (called ``radio''). The ``radio'' of an artist 
is an infinite playlist that includes tracks of this artist and also music by other artists. 
P3 reports that the 
``radio'' based on 
his band includes 
music that he does not like and 
features artists that he distances himself from ideologically. 
\aquote{I see things that I do not like and that I reject ideologically. Why appears [Band~X]?}{P3}
P3 explains that the band works hard on creating 
a certain public image---with the lyrics, the music, the art. The \platform{} 
then mixes it with something different. 
\aquote{I don't think the same people listen to it. That bothers me as an artist.}{P3} 


P1 states 
that the current \platforms{} disconnect the music from its context. 
He points out that a song is inseparable from its social context.
\aquote{Music---art---is a representation of people's sensitivity, it's a diary of 
people telling what happens.}{P1} 
\aquote{Listening to hip-hop from the '90s [is tied to] 
the slums of Los Angeles, of New York, and [to] 
what was happening at that time. That music comes from a place. It doesn't come out of nowhere.}{P1}
P1 thinks that current \platforms{} do not provide much 
context of the music and emphasizes that including such information would 
enhance the experience. 
\aquote{[...] there are songs that have their history, their function. So the more information there is---who are the people who made that song, with whom, how, where, why they sing it---, I know it would be much richer.}{P1}
Yet, P1 adds that some music may not convey any deeper message but exists for business reasons only.
\aquote{[They are] made to sell more records.}{P1}
He puts the example of the genre Reggaeton, for which he thinks that frequently the explicit video is the selling argument and not the music. In such a case, adding information about the context would not add much to the user experience.

\subsection{Reaching an Audience}\label{subsec:reachingaudience}

Another frequently mentioned issue is the difficulty to reach a larger audience, either because the artists are newcomers or when established artists enter a new \platform{}. This issue is usually known in recommenders systems as the item cold-start problem. While it has become easier than ever before to access an enormous amount of music, the artists (P2, PN1, PF1, PF2, PR1) state 
that it is not easy to discover less popular artists with current \platforms{}; it requires the user 
actively looking for those artists when encountering them via other sources such as magazines or interviews. 
PF1 feels that it is very difficult to reach a larger audience. 
\aquote{If [a track] hasn't been listened to a number of times, it does not show up, 
and that means that it is not being recommended.}{PF1}
\aquote{[...] if [the \platform{}] does not recommend things that [...] have rarely been listened to, then it enters a circle that never ends... it always goes... you will never be listened to more. Then you're stuck there until someone pays for you to have the promotion.}{PF2}
P2 also participates in a less-popular band project. He affirms that \platforms{} make it difficult for the users to reach the band. 
\aquote{[...] it would not hurt if the systems were more random, not that obvious---if you like `Beatles,' I recommend `Rolling Stones.'}{P2}

PN1 underlines that the emergence of large \platforms{} changed the entire music industry, making it 
even more difficult for new artists to reach a larger audience. 
\aquote{Before you could go [to a label,] with something super weird but super interesting that could catch their attention and take you on a tour [...]. [But now] it's not the music that sells. [If] you don't have followers, you don't have content, you have nothing, you're nobody, and that's why you won't appear [in the recommendations]. You have to grow in a different way, through Instagram for example, which doesn't have anything to do [with music], and the value of music gets lost.}{PN1}

Also, PR1 points out the difficulty of getting visibility on the \platforms{} if the artist is not popular. Similar to PN1, PR1 argues that it was also hard for an artist to reach visibility before the emergence of \platforms{}.
\aquote{[...] in this business, there have always been many traps. The musician wants to make music, but it's a business. At first, you don't want to see it because you want to make music and you are happy. So for example, the old record companies used to have a monopoly before the [social] networks. You were only played on the radio if they paid for it. They said, `[This artist] has sold twenty thousand copies' but it was the record company itself that bought twenty thousand copies. [...] then everyone wanted to hear that [artist].}{PR1}
\aquote{On YouTube, there were companies that made it to reach the million [...] 
And I think that [happens] today too. You can buy visits [...].}{PR1}
P3 adds that artists being excluded, making it hard to reach an audience, 
has happened ever since, and he provides 
an anecdote of Bob Marley going to a radio station to force them to play his songs. 
\aquote{They had Bob Marley and they didn't play it!}{P3}
P3 argues that 
artists who really want to reach an audience 
will find a way to do it. Yet, probably not through a \platform{} but using other digital mediums. 

In conclusion, there is no clear consensus about the actions that \platforms{} should take 
to be fair in this context. Among the mentioned alternatives, we see that some artists (e.g., P1, P2, P4, PN1) suggest that the \platforms{} should have a minimum quota of starting artists that are recommended to all the users alike. Others suggest 
that new artists should use alternative ways to reach an audience.

\subsection{Transparency}\label{subsec:transparency}

Several artists (P3, PN1, P1, P2, PR1, PF2) mention that the \platforms{} should be more transparent. P3 states that 
he does not understand how exactly the \platform{} promotes some artists more than others; e.g., in the automatic playlists or the curated playlists. \aquote{It would be nice if the platform was equitable, or fair, for everyone in that sense, because if I'm one of the largest bands in Uruguay, why I'm not in many of the playlists there? Is it the platform that doesn't want me there? Is it me 
doing something wrong? [...] Maybe the platform does not benefit much 
with what we do, so they discard us.}{P3} 
PF2 thinks that the \platforms{} should be more transparent towards 
the artists about how their recommendation system works and what the artists have to do for being recommended more often. PF2 thinks that this is particularly important for independent artists. \aquote{[...] you are a bit naked there. You put your music on Spotify and mention in the concerts that they can listen [to your music], but you don't see any change. For example, no one explains to you that it is important that other people add your songs to their playlists so that the algorithm will recommend you.}{PF2}

PJ1 feels that \platforms{} are profitable only for some of the stakeholders.
The artist wonders 
what the goals of the \platforms{} are. 
\aquote{Is the goal for people to listen to music or is the goal to make money from it?}{PJ1} 
\aquote{The people who invest in these things at the [...] 
powerful industry level, these people do make money from this. The others do not. [...] there is no middle class. There is an upper class and a lower class.}{PJ1}

Regarding transparency in algorithmic decisions, the artist PN1 mentions that although both humans and algorithms may be biased, a non-ideal decision made by a human may be easier to accept than one taken by an algorithm. 
\aquote{[...] if a human makes that decision if he says `I have a small store, I am going to put it this way,' I will understand it much more than if an algorithm does it.}{PN1}

\subsection{Influencing Users' Listening Behavior} \label{subsec:influencing}

One of our interests was to learn what the artists think about the \platforms{}' opportunities and power to influence the users' listening behavior. For example, by tailoring the music recommendations, a \platform{} could 
balance 
genres 
or consider a gender balance among the recommended artists. 
In an open question, which did not mention gender, all artists came up with the issue that content by female artists is not well represented. 
We found that all artists agreed that the \platforms{} should promote content by female artists 
to reach a gender balance in what users consume. 

While there was clear consensus to influence users' listening behavior with respect to the artists' gender---to reach a balance---, there was also a clear agreement \emph{not} to do so concerning 
the music style. 
For the latter, they think it is better not to influence the users.
P2 suggests 
a gender balance in the recommendations.
\aquote{[Platforms have] a huge responsibility in making recommendations.
}{P2}
Regarding gender balance, PN1 states, \aquote{[...] the population of the world is 50\% women. So it would be ridiculous if the system wouldn't recommend them.}{PN1} 
PN1 suggests a progressive change, which he thinks will prevent users to perceive it as something negative 
and leave the \platform{}.
PF1 states that the system could enforce a 50\% balance of male and female artists 
because 
many other factors different from gender, e.g., the music style, 
define whether 
someone will like what is recommended.
Finally, PF2 considers that using quotas alone would not be enough, as there is a need for a change in education to have a bigger impact. However, she agrees that artists could be better off with quotas. 
\aquote{As a female artist I would like the system to recommend my music to someone that only listens to the music of male artists.}{PF2}

\subsection{Popularity Bias}

Researchers have put a lot of effort into reducing the popularity bias and improving 
the recommendations for 
items in the long-tail of the popularity distribution (e.g., \cite{vall2019order}). 
Reaching out to the artists, we wanted to explore 
their perspective on that issue; 
whether or not they feel that current systems allow 
the users 
to 
access items of less popular artists; 
and how this could be improved in the future.

P4 states that it may be easier for the \platforms{} 
to recommend what is 
popular because 
it can 
satisfy the majority of the users. Yet, P4 points out, 
\aquote{The problem is [that] the recommender system systematically ignores all those potential artists [...]}{P4}.
P4 thinks that some users may be happy with the recommendation of the generally popular items, whereas others may 
not and probably leave the \platform{}. The users who are more passionate about 
music will not see any advantage in using such 
recommendations.
\aquote{[...] it's wrong that you don't have the option to explore that long-tail. [You can't] take advantage of a recommender system if you want to [explore the long-tail].}{P4}
PR1 considers that the \platforms{} may generate higher revenues 
if they recommend popular artists, and may therefore not be interested in promoting 
less popular content.

Although strongly advocating the promotion of diverse content, PF2 speculates that this may again lead to having users listen to widely popular music. 
Therefore, she claims that 
the \platforms{} should prevent it, \aquote{[...] otherwise, you will always end up listening to American music.}{PF2}


All interviewees 
agree that it is crucial that the \platforms{} also recommend less popular music. They believe that the \platforms{} will harm the music culture if recommendations are limited to the most popular artists. 

\subsection{Artists' Repertoire Size}

While popularity bias is a widely researched topic in the recommender systems community, and for MRS in particular, little attention is paid towards how the size of an artist's repertoire affects the probability of their songs being recommended to users. The artists' opinions are divided concerning how a \platform{} should reflect the differences in the repertoire sizes. 
While three artists think that artists with larger repertoires should be more represented in recommendations (P1, PF1, PR1), four artists do not support that idea (P2, P4, P3, PF2), and two artists were indecisive (PN1, PJ1).

P1 argues that the higher number of records leading to an increased likelihood for an artist's items being recommended reflects what happens outside the \platforms{}.
\aquote{[...] that's fine, the same thing happens in real life if someone makes 25 records, you will surely come across it at some point.}{P1}
PJ1, P2, and P4, in contrast, argue that having more records should not be a reason for being recommended more often.
\aquote{
[...] I know so many amazing bands with only one album, it has 10 songs. And you will never get to those [being recommended].}{P2}
\aquote{This is delicate because there are big artists that have one album, or the opposite.}{PJ1} 
\aquote{Intuitively, I think it is unfair that an artist with more music is recommended more. [...] if an artist has 30~albums but they are all completely different from the previous ones, then it makes sense. But if there is an artist whose 
albums are all the same ...[it does not].}{P4}

PR1 points out that artists with 
more songs are probably more diverse in their repertoire. So, it is more about the diversity than the size of the repertoire. \aquote{If you have more songs you have more chances to satisfy different audiences.}{PR1}

P3 states that an artist profile with more songs may leave the impression that the artist is in the music business for a long time, which may be a reason to recommend the artist more. But he adds that repertoire size should not be given such importance.
PF2 adds that sometimes not all tracks of an artist are available on a particular \platform{}, so the repertoire size on the \platform{} would not reflect reality. 

PN1 raises the issue that it may depend on the users' goals whether the artists' repertoire size matters. 
If a user wants to explore a new artist, then it is not useful to recommend to this user an artist with only a few songs. For users exploring on a track basis, the artists' repertoire size is irrelevant. 



\subsection{Quotas for Local Music}

Today's most prominent \platforms{} operate in several countries, more or less globally. With the widely adopted collaborative filtering approaches for 
recommendations (``people who like that..., also like that...''), it may happen that the music preferences of users in countries with a large number of users 
may influence the algorithms' outputs globally. As a result, artists that are popular in countries with a smaller user number will have fewer chances to be recommended than artists that are popular in countries with a large number of users. 

While there are studies investigating the existence of local trends on global music streaming (e.g., \cite{way2020local}) and recommendation approaches that account for country-specific music preferences (e.g., \cite{bauer2019_PLOSONE}), it is not clear whether and how current \platforms{} consider local repertoire. 
Similarly, outside the \platforms{}, some countries define quotas of local content for radio stations 
while other countries do not. Yet, such quotas for radio do not apply to online \platforms{}, specifically not for automatic recommendations.

We asked the artists 
about quotas, the desirability and applicability to have quotas on online \platforms{}, and for automatic recommendations in particular. We also asked for potential alternative solutions to deal with local content. 
%
Three artists mention that the recommenders should have quotas for local music. Other five artists were not sure whether quotas were the right solution but emphasized that it is important that the \platforms{} promote local content.

PR1 agrees with quotas and indicates that \aquote{otherwise, you won't know what there is in your country.}{PR1}
PJ1 is ambivalent. 
On the one hand, he believes that it is not right to base such decisions on where a person is from, 
on the other hand, he sees the need to promote local artists because they are at a disadvantage from the start; and with quotas, they could make up for it over time.

P1 suggests abandoning the idea of defining locality in terms of countries because national borders are not necessarily cultural borders. He proposes to define locality within a radius of a user: Artists that are in a certain radius (e.g., within a radius of 5000~km) should be given a higher weight than artists outside that radius; and within the radius, different weights again, with higher weights for the closest artists. 
Yet, the same artist (P1) emphasizes 
that quotas are a necessary measure in some countries because, otherwise, local artists would not be able to make a living solely from music. Accordingly, he suggests that quotas should also apply to automatic recommendations and proposes to use 
a combination of country and radius.
\aquote{[...] if you go to the border [between Uruguay and Brazil] [...], the Brazilian influence is greater than the Uruguayan one. So it seems to me that the radius is more representative for culture.}{P1} 

PN1 considers it peculiar that 
there is a higher chance for the user to be presented with US artists compared to local ones, even if the latter are locally famous and popular. PN1 calls for more transparency and draws
an analogy to the news sector. 
\aquote{[...] it is like reading the news in the New York Times instead of the local newspaper. [...] 
you know that you're reading the New York Times or the local news. But you don't know 
whether it is an algorithm that makes the recommendation.}{PN1}

PF2 and P4 argue against quotas because this 
could cause users to leave the \platform{} if they do not like local music. Yet, both emphasize giving 
importance to locality. P4 suggests giving individual users the chance to choose the degree of locality they want to have. 
PF2 suggests promoting 
local content by letting users indicate the countries they 
would like to receive recommendations from besides their own country. Artists could also be allowed to indicate in which countries they would like their music to be recommended. 
This would enable artists to reach other countries. 
\aquote{As an artist, you could reach more countries if you are interested.}{PF2}

P2 is unsure whether quotas are the ideal measure but emphasizes that the \platforms{} have responsibility for what their algorithms recommend. 
\aquote{[A platform provider] needs to take responsibility for its recommender system---[...] understand the situation. [...] Obviously, they are commercially not capable or not interested [...] but it would be great [...] if [they] find a way to link [...] a Yankee band with a Uruguayan band [...] 
make a connection that contributes. [...] [The platforms] have their share of responsibility for what they are showing or recommending. I don't know if there should be quotas [...] but [...] it would be great if there was something.}{P2}

P3 questions whether quotas should be enforced by law and suggests that \platforms{} take the responsibility for it.
\aquote{The question is whether it should indeed be state-imposed. For things to be that way, do we have to impose it?}{P3} 
\aquote{[the platform] should do what is ethically correct. [...] If I were Spotify [...] I would [promote local content] in every country.}{P3}
While P3 voices concerns about whether the \platforms{} should be trusted in deciding 
what to promote more or what to promote less. Yet, for gender fairness and local content, P3 is confident that the \platforms{} could find the right balance.

Different from the common understanding of local quotas, PF1 suggests that it should be the opposite: Instead of having quotas for local music in smaller markets, there should be quotas in larger markets to include music from those smaller markets.
In addition, PF1 points out that giving users the possibility to 
explore the country-specific music scene would be 
more beneficial for the artists. 
\aquote{[...] provide the possibility to listen to what you have not listened to before. For example, `what have I not listened to from Colombia?' `What is underground in such a country?' If you give the local artists a voice and let them tell the story behind their music, that would be more interesting.}{PF1}

\subsection{New Music}\label{subsec:newmusic}

New music may refer to (i)~artists that are new to a user (thus, the discovery of artists) and also to (ii)~a new track or album released by an artist that had already been part of the \platform{}.

Most of the interviewed artists agree that the artists should be in control of what tracks or albums get more recommended. In the case that they are not in control themselves, they strongly prefer a recommender system that puts more weight on their latest releases. PR1 states that every artist wants their new music to be promoted so that the world finds out they have a new release.
\aquote{[...] you do a promotion campaign to tell that you released more [content]. To tell the world, `Hey! There is a new album!' [...] Like saying, `Hello, I'm here.'}{PR1}
Also, some artists feel more identified with what they are doing now compared to music they released many years ago, which is another reason for them to prefer the promotion of the latest release. 
\aquote{[...] it is something that I have done 10~years ago. [...] I don't know if I feel identified [with it].}{P3} 
 
For allowing users to discover artists that they are not yet familiar with, there is no clear consent either. While all agree---in varying degrees, though---that the \platforms{} should allow users to discover artists that they do not know, it remains unclear how the \platforms{} should do that.
Most of the interviewed artists state that the user should be in control, having the opportunity to indicate that they want to discover new artists, and when they want to do so. 
\aquote{[...] being able to choose would be good. A button that says `I'm open to new stuff' or `Let me listen to what I want.' Because sometimes you want new stuff and sometimes you want something very specific.}{P2}

\section{Discussion}\label{sec:discussion}

Table~\ref{tab:concl} summarizes the concrete aspects 
in which the \platforms{} could be more beneficial for the artists. The results show that the artists' perspective on fairness relates closely to the end consumer's perspective on some aspects, whereas the interviews could also reveal that artists face problems that are not reflected in previous work on the consumer's perspective. The different roles (here, users vs item provider) may lead to varying perceptions of fairness (see~\cite{wang2020}). For example, gender fairness 
is frequently also discussed from the consumer or societal perspective~\cite{sonboli2021fairness}. Our results make concrete that artists aim for \emph{gender balance}. 
In contrast, the problems associated with the fragmented presentation (e.g., music presented detached from its context; artist profile may list tracks first that am artist does not identify with anymore), for example, accrue from the artist perspective. Furthermore, the results suggest that not all identified aspects that artists consider important for a fair \platform{} are directly linked to algorithmic fairness. For instance, the perceived lack of control and the demand that music is presented in a way so that its context is clear are not of algorithmic nature; rather, these are system design issues. Hence, the HCI community is called to address the needs of this so-far underrepresented group.

\begin{table*}[hb]
  \centering
  \caption{Aspects to improve with most of the artists in agreement.}
  \label{tab:concl}
  \begin{tabularx}{\textwidth}{l X} 
  \toprule
  Topic & Description\\
  \midrule
   Quotas for local music&Promoting local music\\
   Gender balance & Expectation of gender balance in recommendations\\
   Popularity bias & Recommending items in the long tail, 
   not only the most popular artists\\

   Lack of control & Giving artists control concerning the tracks that are promoted; if not in control, preference for 
   promotion of 
   latest releases\\

   Transparency & Transparency 
   about how the algorithms work; why is music 
   recommended or not \\
   
   Influencing users' 
   taste & The system should not influence the user's taste \\
   Music in 
   context & 
   Music should be presented to users with information about its 
   context\\
  \bottomrule
  \end{tabularx}
\end{table*}

%

For operationalization, 
we can build on the strong foundations of prior research. For instance, from the interviews, we understand that artists see the need to promote new and less popular artists. While collaborative filtering is commonly used in MRS, it is an approach that is prone to popularity bias. 
Content-based approaches based on the advancements in music information retrieval 
(see~\cite{Murthy2018}) 
could be especially apt to promote new and less popular artists.

Furthermore, as different the topics of promoting local music and ensuring gender balance in recommendations seem the basis for their operationalization exhibit similarities. First, meta-data about both, the artists' regional or cultural affiliation as well as gender information, are available for popular artists (e.g., using sources such as Wikipedia or MusicBrainz), but scarce for new and less popular artists. Thus, while existing meta-data may be used, other approaches have to be leveraged to gather missing data; this may be challenging for new and less popular artists, in particular. Second, 
many 
works investigate 
the diversity or coverage of computed recommendations 
(see~\cite{KUNAVER2017154}), little is known about how to ensure an envisaged ratio of attributes (here: region, culture, gender). In~\cite{ferraro2021_break}, we proposed a re-ranking approach to gradually achieve gender balance. Similar approaches may be used for other attributes. Targeting ratios for multiple criteria is more complex.

In addition, the finding that artists perceive their profile presentation 
as being fragmented 
and that their music is presented detached from its context, 
are an inspiration and rationale 
to consolidate and structure information from dispersed sources, so that the music presentation can be enriched with this information and put into context. 
For new and less popular artists, but also for new music by established artists, it will be challenging to retrieve such information. Besides challenges concerning the operationalization for information retrieval and consolidation, it is subject to future research to investigate how such contextualized information should be presented so that it (i)~puts the music into context as 
meant by the 
artist and (ii)~is understandable and appealing to the user.
This is a great opportunity to also address the artists' demand for more control and to add interaction functionality for this stakeholder. This functionality may be as simple as an interface to insert context information or correct information that was 
retrieved from other sources. Providing templates for visualizing context information may add control. Further, templates and functionalities to give the artist page a personal branding may be appreciated by artists because the visual representation has long been a strong component in the music field (e.g., album artwork, stage shows, the association of genres with colors, etc.).

\section{Conclusions}\label{sec:conclusion}

We reached out to music artists 
and conducted 
semi-structured interviews 
to understand how they are affected by current \platforms{} and what improvements they deem necessary so that 
those \platforms{} are 
\emph{fair} from the artists' perspective. Thereby, we paid particular attention to music recommender systems that are an integral part of today's \platforms{}. We conclude that the participating artists' perceptions and ideas are well aligned. 

The interviewed artists agree on some aspects that are needed to make the \platforms{} fairer: 
1)~The artists call for better promotion of local music; 2)~they 
agree that gender balance in the recommendations is indisputably 
expected; 
3)~the artists voice that music items in the long tail of the popularity distribution (not only the most popular artists) have to be included in the recommendations shown to users; 
4)~the participants advocate giving 
control to the artists 
over the 
tracks that are 
promoted or higher weighted in recommendations (if they are not directly in control, then they generally favor the promotion of their latest releases); 5)~they request transparency about how the algorithms work, to understand why their music is recommended or not; 6)~the artists consider a system that influences a user's taste (or attempts to do so) an undesired misuse; 
and 7)~they would appreciate if the \platforms{} would be enriched with information that puts the music into context. 

Besides the 
consensus 
on many topics, 
there is no clear direction for others: 1)~There is no agreement concerning 
quotas 
for local content; 2)~no clear majority 
whether the size of the artists' repertoire should be reflected in recommendations; 
3)~while the artists seem to agree that new 
artists should be given space on the \platforms{}, there is no agreement on how to operationalize this; 
and there is 4)~no clear agreement whether a \platform{} should promote the discovery of artists previously unknown by a user. 
%
Overall, while there is a prevailing belief that 
\platforms{} 
would open up 
the long-tail to users 
and encourage them to consume more of those items, 
the interviews suggest that 
the long-tail items and artists remain obscure. 

A limitation of our work relates to the chosen context. We deliberately 
drew the sample from a big and coherent music market (Spanish-speaking music market), which is geographically dispersed (Europe, North America, and South America), and where the artists are diverse in styles, career, and popularity. While we could find shared ideas and opinions across the sample which is diverse in many aspects, the sample's perception could still be aligned from a cultural perspective (e.g., Uruguay and Spain are similar in most of Hofstede's cultural dimensions~\cite{hofstede2010cultures}). Yet, this defined setting (artists from Spanish-speaking countries) is a good opportunity for upcoming research to draw comparisons to other defined groups of artists.
%
Future research may 
investigate the identified aspects in different contexts and more depth. Further, while 
we catered for some aspects of diversity in our sample of music artists, reaching out to a larger sample will allow for even more aspects of diversity (e.g., 
including artist of non-binary gender, considering solo artists, mixed-gender bands, mono-gender bands, various ethnic groups, artists dedicated to 
niche music). 
Although we reached thematic saturation, 
reaching out to a wider set of diverse artists may reveal additional 
topics or different viewpoints.

These caveats notwithstanding, our work gives direction towards relevant topics for fairness on \platforms{} and their integrated MRS. The findings indicate pathways towards fairer \platforms{}, whereas the concrete operationalization is subject to further research. 

Naturally, the music domain has its specificities. Yet, our study results may inspire studies in other domains. 
First, our findings indicate that different roles may come with different fairness perceptions and requirements. Hence, we encourage 
studies in other domains where 
an item provider typically has several items and where a person is the face to the public for themselves, an organization, or a brand (e.g., sports, many technology companies).
Second, our findings suggest that perceived unfairness also is tied to a lack of control. Likely, control may also be given 
with more opportunities for interaction.




\section*{Acknowledgments}
This research was partially supported by Kakao Corp.

%
%
%
 \bibliographystyle{splncs04}
 \bibliography{artists_bib}
\end{document}